\documentclass[12pt]{article}
\usepackage[left=3cm,right=3cm,top=3cm,bottom=3cm]{geometry}
\usepackage{fancyhdr}
\author{George Hassan-Coring}
\date{\today}
\usepackage{cite}
\usepackage{bm}
\usepackage{graphicx}
\usepackage[normalem]{ulem}

\begin{document}
\begin{center}
\title =  \textbf{{\LARGE Inferring Causality in Agent-Based Simulations - Literature Review}\\George Hassan-Coring\\Department of Computer Science\\University College London\\December 2018}
\end{center}


\begin{abstract}
Complex systems have interested researchers across a broad range of fields for many years and as computing has become more accesible and feasible, it is now possible to simulate aspects of these systems. A major point of research is how emergent behaviour arises and the underlying causes of it. This paper aims to discuss and compare different methods of identifying causal links between agents in such systems in order to gain further understanding of the structure.
\end{abstract}

\section{\underline{Introduction}}
\par The statistical analysis of complex systems has interested academics for the last sixty years and this is not only due to the difficulty of the problem but also the wide range of possible applications of it. Complex systems can vary between biological ecosystems \cite{KBentley2012,KBentley2004,KBentley2005,CCChen2007c,CCChen2008a} to social media interactions\cite{social1,social2}, financial markets \cite{TChiotis2007,PAnderson,CHiemstra} to weather prediction\cite{Wiener1956}. This range of possibilities for complex systems means that researchers seek to find some common ground that link these systems which can help in understanding them as a whole and how these systems develop over time. Computer simulations are an essential tool \cite{CClack2006,CCChen2007a,CCChen2007b} in this investigation as they allow repeated analysis in great detail. This literature review first provides basic definitions and then presents an overview of relevant work in this area.\\

\section{\underline{Definitions}}

\par A \textbf{complex system} is ``one in which there are multiple interactions between many
different components''\cite{Science1999}. The \textbf{time series} of output values from these components are values measured at each discrete unit of time. The time series are influenced by the interactions between components and we are interested in discovering causal links between these time series. An example of this could be the price of an individual company's stock over a time period and how it is linked to interest rates. A related property is the \textbf{stationarity} of a time series where ``the stochastic mechanism generating the sequence is not changing'' \cite{stationary} and so the parameters like the mean or variance do not change over time. This property allows us to properly use statistical tests and techniques on the time series, making it very important. \\

\par \textbf{Causality}, as mentioned in this paper, will generally refer to \textbf{Granger Causality} \cite{Granger1969} and not \textbf{True Causality} \cite{Leo} unless stated otherwise. True Causality is the underlying relationship between causally linked variables and discovering it is the ultimate goal of investigating causality. However, problems arise from things like random noise and error in measurements leading attempts to find true causal links astray so alternative methods are needed in practice.
 Granger Causality is a statistically testable measure of causality applied to time series in a system. To determine whether one variable $X$ `causes' another variable $Y$, a hypothesis test is used to compare whether a prediction model using the time series of $X$ up to that moment as well as the time series of $Y$ prior to that moment is better at predicting the next value of $Y$ than an alternative prediction model using just the time series of $Y$. \\

A related term is \textbf{Spurious Causality};  this is ``incorrect causality, in a multivariate system, due to common drivers or indirect interactions.''  \cite{Leo}  Finding these spurious links is important because it allows us to get a better picture of the whole system without these unnecessary links polluting it.\\

\par When discussing complex systems, a major point of inquiry is \textbf{Emergent Behaviour}, defined as 
``A property is emergent if and only if it is present in a macrostate and it is not present in the microstate." \cite{ARyan} These emergent properties are so interesting because they are built incrementally by the simple definitions and interactions of multiple agents across an entire complex system and are the subject of a great deal of research. \cite{CCChen2007a,CCChen2007b,CCChen2008a,CCChen2008b,CCChen2009a,CCChen2009b,CCChen2010} A simple example could be an upper limit for a value in a time series that is not explicitly defined but which emerges over the time span of the system. Understanding the causal links of a system is the first step to investigating and understanding the emergent properties.
\\\\



\section{\underline{Literature Review}}

\par The first paper that truly initiated research into the analysis of complex systems was Weiner's paper in 1956 \cite{Wiener1956}. In this piece of work, the context was meteorology and the prediction of weather systems. He discusses what makes the prediction of systems like this so difficult like the error of measurements and how these errors can propagate in deeply connected systems such that our predictions are affected. His work began to introduce statistical methods into the field so that it could be formally reasoned about and inspired others to build on his work. The impressive thing about this paper is that his comments on common errors that arise are just as relevant to this paper today as they were sixty years ago. \\

\par Inspired by Wiener's previous work, Granger went on to publish his hugely important paper in 1969 \cite{Granger1969}. In this paper, he outlined the method which can be used to determine a statistical cause and when this method can be used, giving rise to the term Granger Causality as defined above. Differing to Wiener, Granger discussed these methods to be used in the field of econometrics, showing that even this early just how diverse the applications were. This paper is a true cornerstone of the whole subject, being cited over 20000 times according to JSTOR and his methodology is favoured today in fields like econometrics and neuroscience. The simplicity of the method allows it to be easily applied and is one of the reasons it has been chosen for my own work concerning the Interdyne simulator. \\

\par In Granger's 1969 paper \cite{Granger1969}, he begins by explaining how a stationary time series can be ``decomposed into unrelated components associated with a particular frequency", where stationary time series refers to a time series whose statistical properties like mean and variance stay constant over time. The variance of this time series is the sum of the variance of these individual component time series. \\\\


\par Granger also mentions some interesting situations that can occur like spurious causality in the trivariate case (mentioned previously) and instantaneous causality \cite{Granger1969}. Instantaneous causality describes a causal relationship between variables $X$ and $Y$ for example, where $X$ is best predicted at a moment in time, $t$, using the previous values of $X$ before point $t$ as well as the values of $Y$ before and \textbf{including} the value at time $t$. The opposite of this would be simple causality where $X$ depends only on $X$ and $Y$ values previously to time $t$. A fascinating point that Granger makes about these possibilities is that a relationship may appear to be instantaneous causality but this may be due to the frequency of the measurements of the data rather than the actual properties of the variables. For example, instantaneous causality is detected when using data measured every 2 seconds but in fact there is a time lag of one second between the variables that is missed due to the infrequent sampling.\\\\

\par Granger aims to investigate causal feedback in systems, something which he feels was not pursued prior to his paper so he sets out some formal definitions for causality and feedback which provide the basis of the Granger test.
\begin{itemize}
\item \textbf{Causality:} if $\sigma^2(X|\bar{X},\bar{Y}) < \sigma^2(X|\bar{X})$ then $Y \rightarrow X$
\item \textbf{Feedback:} $X \leftrightarrow Y$
\end{itemize}
His definition of causality allows him to model possible distributions for variable $X$ and evaluate them using the variance in order to check which one provides the best estimate and thus, whether $Y$ has a causal relationship with $X$. This is the essence of the Granger test for causality but the test statistic can be altered from variance of the model to different values. Granger himself acknowledged the limitation of using the variance to judge the models and suggested further investigation. He also mentioned that this theory only applies to stationary time series as non-stationary series are much harder to create tests for. Examples of bivariate and trivariate tests are presented in the paper.\\\\


\par An outline of the process of determining Granger Causality between two variables, $X$ and $Y$, would be first to generate two possible models for the variable $Y$ at a time $t$:
\begin{equation} \label{eqn1}
 Y_{t+1} = \sum_{i=1}^{t} \alpha_{i} Y_{t-i} + \epsilon_t
\end{equation}
\begin{equation} \label{eqn2}
 Y_{t+1} = \sum_{i=1}^{t} \alpha_{i} Y_{t-i} + \sum_{j=1}^{t} \beta_{j} X_{t-j} + \epsilon_t
\end{equation}
\par These models provide the null and alternative hypothesis respectively for our hypothesis test. Typically a Wald Test is used to determine whether the null hypothesis is rejected or not as it is used to determine whether exploratory variables are significant with respect to a test variable. According to Aggresti \cite{wald}, Wald tests are generally more widely applicable compared to alternatives like the Likelihood Ratio Test or Lagrange Multiplier Test as Wald tests require less knowledge about the distribution of the time series. If the Wald test finds variable $X$ to have statistical significance in predicting variable $Y$, we reject the null hypothesis (\ref{eqn1}) and accept the alternative hypothesis (\ref{eqn2}). Thus suggesting that $X$ ``Granger causes'' $Y$.\\\\

\par Further papers in 1980 \cite{Granger1980} and 1988 \cite{Granger1988} further discuss the limitations and criticisms of Granger's methods which range from philosophical considerations of whether it's appropriate to use the word ``cause" in relation to these variables to whether defining instantaneous causality is necessary. Granger et al \cite{Granger2000} give a good summary of his methods, and uses Granger causality tests extended with further econometric measurements to analyse causality between Asian stock markets and exchange rates. The authors use advanced statistical techniques along with observations about the context of the situation to address the non-stationary property of the variables in this time period.\\\\

\par An alternative way of investigating causality between time series is proposed by Holland \cite{Holland1986}. Inspired by the work of Rubin \cite{Rubin1974}, Holland was interested in ``measuring the effects of causes rather than the causes of effects" and used the idea of experiments to consider causality. To test whether a variable $X$ causes a variable $Z$, let $Z$ be the response value; this is the variable that is measured in order to determine if the other variables have a causal influence on it. $X$ and $Y$ are the experiment and control values respectively and are the two possible causes of $Z$. He proposed looking at the difference in the response variable after being exposed to the experiment value and the control value to determine causal inference. Holland splits the response variable $Z$ in two to represent the two potential responses from each of the causes: $Z_X$ for the response caused by $X$ and $Z_Y$ for the response caused by $Y$. This differs from Granger's approach in that it always considers causation to be relative, either $Z$ is caused by the experiment value, $X$, or it is caused by control variable, $Y$, and there are no other causes outside of the tester's control.  This approach has its merits when conducting a carefully controlled experiment, for example: a medical experiment, but becomes harder in a less controllable situation. There is also a chance of $X$ and $Y$ themselves being causally linked such that Holland's method would provide no relevant results as $Z_X$ and $Z_Y$ would have similar values. In the context of the Interdyne simulator with many different agents and rules, this method might not produce as many valuable results as Granger's methods because a large number of variables can be present and there could be spurious links present, resulting in many pointless tests using Holland's method. Care needs to be taken with the testing method to ensure that complex cases such as spurious causality are considered as they can easily arise within the system and negatively affect the results of the tests.\\\\


\par Granger's simple testing method has also been extended by others in order to be usefully applied in other fields. For example, Geweke \cite{Geweke1982} presented a method to decompose the causal relationships at certain frequencies, similar to what was suggested by Granger \cite{Granger1969}. This method of decomposition was of great interest to neuroscientists with regards to signals in the brain over time and means that Granger testing can be applied to this field too \cite{Kaminski2001}. Granger's theories are still an active area of research 50 years later \cite{TAste,Leo} and this shows the quality of his work and the developments of others who have built upon it.\\\\

\section{\underline{Conclusion}}

\par In conclusion, Granger causality tests appear to be the best choice for trying to identify causal links between time series as they are hugely popular across a range of disciplines concerning the study of complex systems and are relatively easy to implement. Having investigated some of the alternate ideas to Granger causality, they seem to be more concerned with problems regarding the specific definition of causality and are less practical when implemented. After consideration of the related literature, I believe that Granger causality is the most appropriate choice to evaluate causal relationships in a complex system like the Interdyne simulator as it is more adaptable and easier to implement than the alternatives.

\bibliography{literature}
\bibliographystyle{plain}

\end{document}